\documentclass[aps,prd,eqsecnum,preprint,tightenlines,nofootinbib,showpacs]{revtex4-1}
\usepackage{graphicx,amsmath,latexsym}

\def\bea{\begin{eqnarray}}
\def\eea{\end{eqnarray}}
\def\be{\begin{equation}}
\def\ee{\end{equation}}
\newcommand{\ub}[1]{\underline{#1}}

\begin{document}

\title{A brief introduction to nonperturbative calculations \\
in light-front QED%
\footnote{Contributed to the proceedings of QCD@Work 2010, the international
workshop on QCD Theory and Experiment, Martina Franca, Italy, June 20-23, 2010.}}


\author{Sophia S. Chabysheva}
\affiliation{Department of Physics \\
University of Minnesota-Duluth \\
Duluth, Minnesota 55812}

\date{\today}

\begin{abstract}
A nonperturbative method for the solution of quantum field theories
is described in the context of quantum electrodynamics
and applied to the calculation of the electron's anomalous
magnetic moment.  The method is based on 
light-front quantization and Pauli--Villars
regularization.  The application to light-front QED
is intended as a test of the methods in a gauge theory,
as a precursor to possible methods for the nonperturbative
solution of quantum chromodynamics.  The electron state
is truncated to include at most two photons and no positrons
in the Fock basis, and the wave functions of the dressed
state are used to compute the electrons's anomalous
magnetic moment.  A choice of regularization that preserves
the chiral symmetry of the massless limit is critical for the
success of the calculation.
\end{abstract}


\maketitle


\section{Introduction}

The purpose of this work
is to explore a nonperturbative method that
can be used to solve for the bound states of quantum field theories,
in particular QCD.  The problem is notoriously difficult,
and there are only a few approaches.  These include
lattice gauge theory~\cite{Lattice}, 
the transverse lattice~\cite{TransLattice},
Dyson--Schwinger equations~\cite{DSE},
Bethe--Salpeter equation,
similarity transformations combined with construction of
effective fields~\cite{Glazek},
light-front Hamiltonians with
either standard~\cite{bhm} or sector-dependent
parameterizations~\cite{SectorDependent,Karmanov,Vary}.

We use the light-front Hamiltonian approach with Pauli--Villars 
(PV)~\cite{PauliVillars} regularizaton and 
standard parameterization, where the bare
parameters of the Lagrangian do not depend
on the Fock sector.  This means that we use Fock states --
the states with definite particle number and definite momentum
for each particle -- 
as the basis for the expansion of eigenstates.  The coefficients
in such an expansion are the wave functions for each possible set
of constituent particles.  These functions describe the distribution
of internal momentum among the constituents.  Such an expansion is
infinite, and we truncate the expansion to have a calculation of
finite size.

The wave functions are determined by a coupled set of integral
equations which are obtained from the bound-state eigenvalue
problem of the theory.  Each bound state is an eigenstate of
the field-theoretic Hamiltonian, and projections of this
eigenproblem onto individual Fock states yields these coupled
equations.  Each equation is a relativistic analog of the 
momentum-space Schr\"odinger equation, but with terms that
couple the equation to other wave functions that represent
different sets of constituents, perhaps one gluon more or less
or a quark-antiquark pair in place of a gluon or vice-versa.

The solution of such equations, in general, requires numerical
techniques.  The equations are converted to a matrix 
eigenvalue problem by some discretization of the integrals
or by a function expansion for the wave functions.  
The matrix is usually large
and not diagonalizable by standard techniques; instead, one
or some of the eigenvalues and eigenvectors are extracted by 
the iterative Lanczos process.
The eigenvector of the matrix yields the wave functions, and from
these can be calculated the properties of the eigenstate, by
considering expectation values of physical observables.

We work with light-cone coordinates~\cite{Dirac,DLCQreview},
chosen in order to have 
well-defined Fock-state expansions and a simple vacuum.
The time coordinate is $x^+=t+z$ and the space
coordinates are $\underline{x}=(x^-,\vec{x}_\perp)$, with 
$x^-\equiv t-z$ and $\vec{x}_\perp=(x,y)$.  The light-cone
energy is $p^-=E-p_z$, and the three-momentum is
$\underline{p}=(p^+,\vec{p}_\perp)$, with
$p^+\equiv E+p_z$ and $\vec{p}_\perp=(p_x,p_y)$.
The mass-shell condition
$p^2=m^2$ becomes $p^-=\frac{m^2+p_\perp^2}{p^+}$.
The simple vacuum follows from the positivity of the
plus component of the momentum:
$p^+\equiv \sqrt{m^2+p_z^2+p_\perp^2}+p_z>0$.

To regulate a theory, we use the Pauli--Villars technique~\cite{PauliVillars}.           
The basic idea is to subtract from each integral a contribution
of the same form but of a PV particle with a much larger mass.
This can be done by adding negative metric particles to the
Lagrangian.  For example, for free scalars a Lagrangian of
the form
            \be {\cal L}=
                 \left[\frac12(\partial_\mu\phi_0)^2-\frac12\mu_0^2\phi_0^2\right]
                 -\left[\frac12(\partial_\mu\phi_1)^2-\frac12\mu_1^2\phi_1^2\right] 
                 \ee
generates a contribution from an internal line of a Feynman diagram
in the form
            \be 
            \int \left[\frac{1}{p^2-\mu_0^2}-\frac{1}{p^2-\mu_1^2}\right] d^4p,
            \ee
which has the necessary subtraction.  A particular advantage of
PV regularization is preservation of at least some symmetries;
in particular, it is automatically relativistically covariant.

\section{Application to QED}

The method is not mature enough to apply to QCD, so
as a test in a gauge
theory, we consider light-front QED and
specifically the eigenstate of the dressed
electron and its anomalous moment~\cite{ChiralLimit,Thesis,SecDep,TwoPhotonQED}.
From the PV regulated light-front QED Lagrangian, we construct
the Hamiltonian ${\cal P}^-$ and solve the mass eigenvalue
problem ${\cal P}^-|\ub{P}\rangle=\frac{M^2}{P^+}|\ub{P}\rangle$
in the approximation that the electron eigenstate is a truncated
Fock-state expansion with at most two photons and no positrons~\cite{TwoPhotonQED}.
From this approximate eigenstate, we compute the anomalous magnetic
moment from the spin-flip matrix element
of the electromagnetic current $J^+$~\cite{BrodskyDrell}.
            
\begin{figure}[t]
\centerline{\includegraphics[width=15cm]{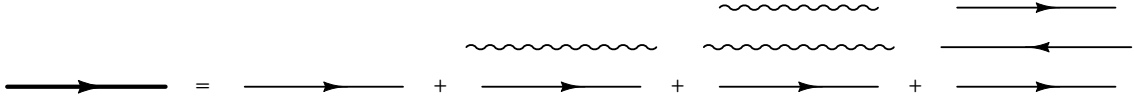}}
\caption{\label{fig:dressedelectron}The Fock-state expansion
of the dressed-electron eigenstate.}
\end{figure}

\begin{figure}[ht]
\centerline{\includegraphics[width=13cm]{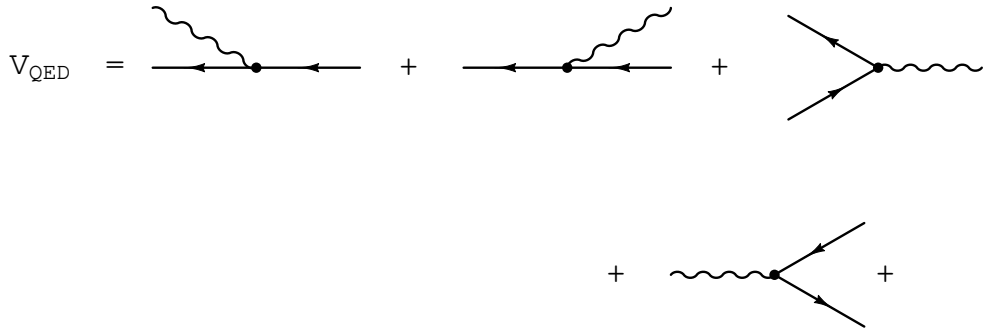}}
\caption{\label{fig:Vqed} The potential-energy operator of light-front
QED in a graphical representation.}
\end{figure}

\begin{figure}[ht]
\centerline{\includegraphics[width=10cm]{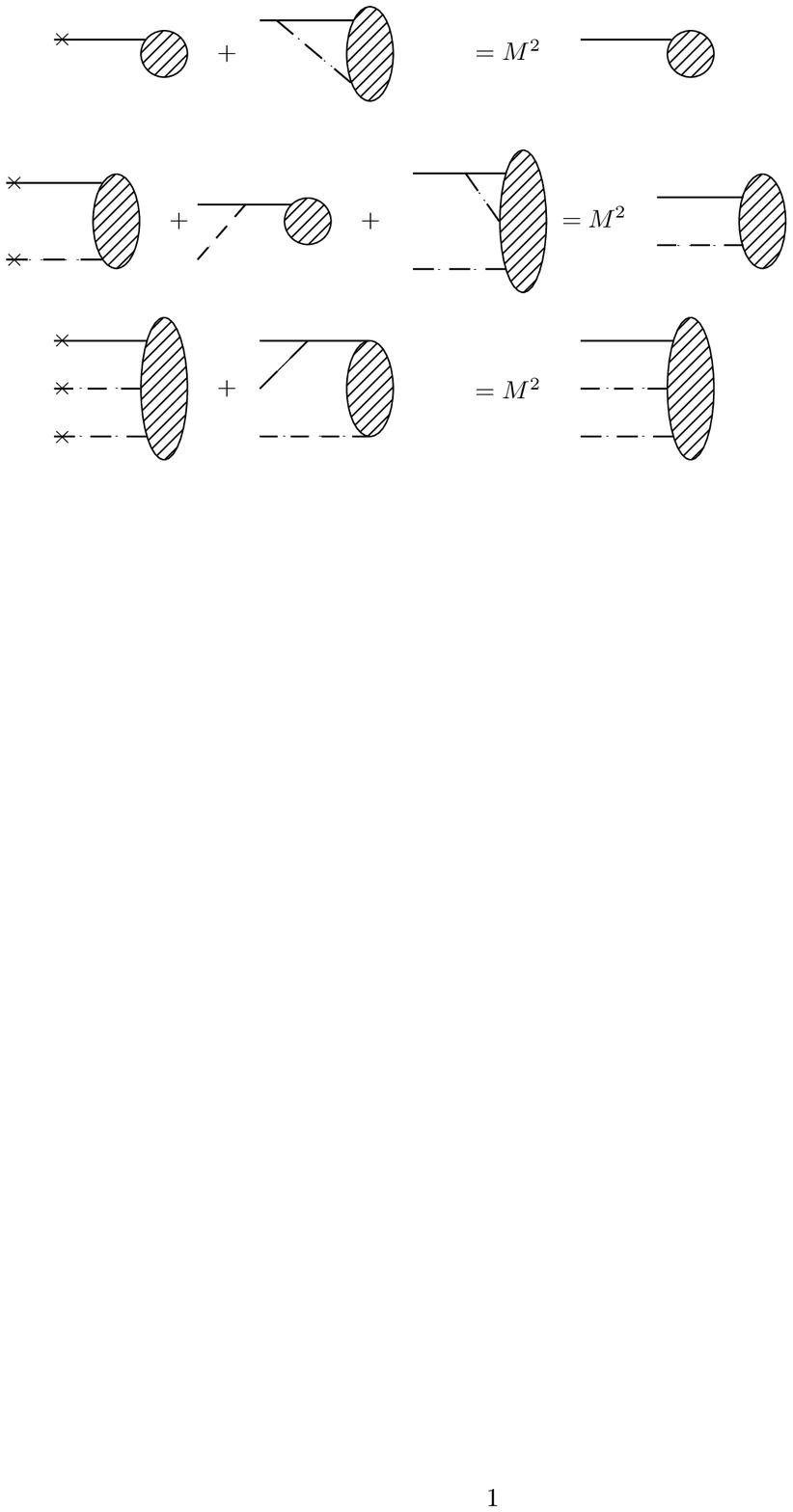}}
\caption{\label{fig:coupledequations} A graphical representation
of the coupled equations for the Fock-state wave functions of
the dressed electron eigenstate.}
\end{figure}

Schematically, the electron Fock-state expansion
can be written
\be
|{\rm electron}\rangle=\int\psi_{e}|e\rangle +\int\psi_{e\gamma}|e\gamma\rangle
     +\int\psi_{e\gamma\gamma}|e\gamma\gamma\rangle
        +\int\psi_{eee^+}|ee^+\rangle+\cdots
\ee
This is represented graphically in Fig.~\ref{fig:dressedelectron}
It satisfies the eigenvalue problem
\be
H_{\rm LC}|{\rm electron}\rangle
         =\left(K+V_{\rm QED}\right)|{\rm electron}\rangle=M^2|{\rm electron},
\ee
where $V_{\rm QED}$ is the potential-energy operator, represented
graphically in Fig.~\ref{fig:Vqed}.
The projections of this eigenvalue problem onto each individual
Fock state produces coupled equations for the Fock-state wave functions,
which are schematically represented in Fig.~\ref{fig:coupledequations}.      
The first graphical equation in Fig.~\ref{fig:coupledequations}
is a projection onto the one-electron Fock state, which
can be written as
\be
m_0^2\psi_{e}
    + \int d\ub{k}_\gamma V_{e\gamma\rightarrow e}(\ub{k}_\gamma)
                               \psi_{e\gamma}(\ub{k}_\gamma) =M^2 \psi_{e}.
\ee
It includes absorption of the photon from the one-electron/one-photon
state.  The second equation can be expressed as
\be
\sum_i \frac{m_i^2+k_i^2}{k_i^+/p^+} \psi_{e\gamma}
      + \int d\ub{k}_\gamma V_{e\gamma\rightarrow e}(\ub{k}_\gamma)
                      \psi_{e\gamma\gamma}(\ub{k}_\gamma)=M^2 \psi_{e\gamma} .
\ee
This includes photon emission by the bare electron and
photon absorption from the one-electron/two-photon state.
The third equation is the analogous one for the three-body sector.

\begin{figure}[ht]
\centerline{\includegraphics[width=10cm]{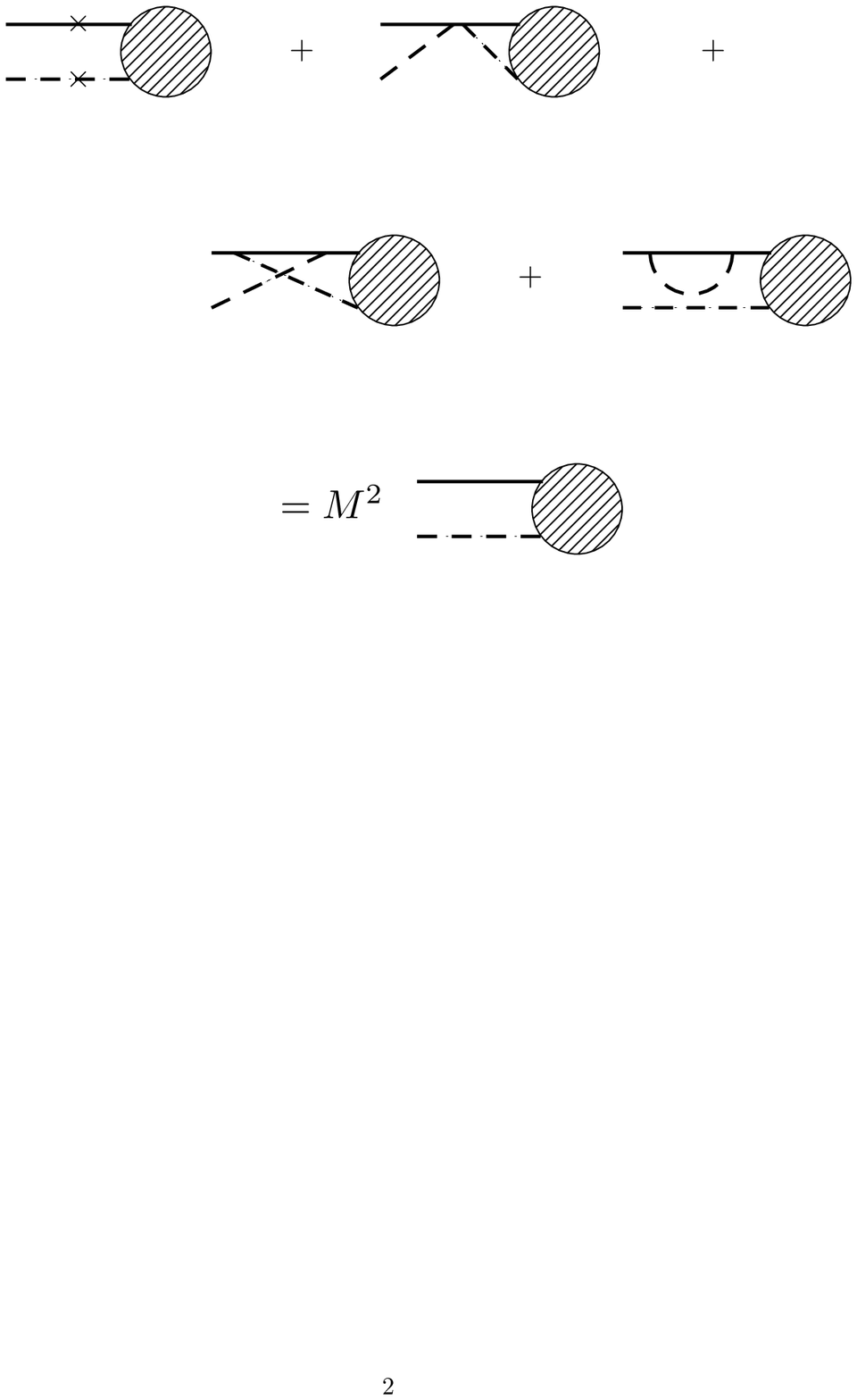}}
\caption{\label{fig:viscacha2} Diagrammatic representation of
effective integral equation in the one-electron/one-photon
sector.The first term represents the kinetic energy; the second
and third, the two time orderings of photon absorption and
emission; and the fourth, the self-energy contribution.}
\end{figure}

The first and third equations of the coupled system
can be solved for the bare-electron
amplitudes and one-electron/two-photon wave functions,
respectively, in terms of the one-electron/one-photon
wave functions.  Substitution of these solutions into
the second integral equation yields a reduced integral
eigenvalue problem in the one-electron/one-photon sector.
A diagrammatic representation is given in Fig.~\ref{fig:viscacha2}.%

Solution of the resulting integral equations
yields $\alpha$ as a function of $m_0$ and the PV masses.  Then for given
values of PV masses, we can seek the value of $m_0$ for which
$\alpha$ takes the standard physical value $e^2/4\pi$. 
The equations must first be solved for $M=0$, with the 
coupling strength parameters adjusted to yield $m_0=0$~\cite{TwoPhotonQED}.  
The solution of the integral equations requires numerical
techniques~\cite{TwoPhotonQED}.  The integrals are discretized
via quadrature rules, and the equations are thereby converted
to a matrix eigenvalue problem, which is solved by iteration.

\begin{figure}[tp]
\centerline{\includegraphics[width=14cm]{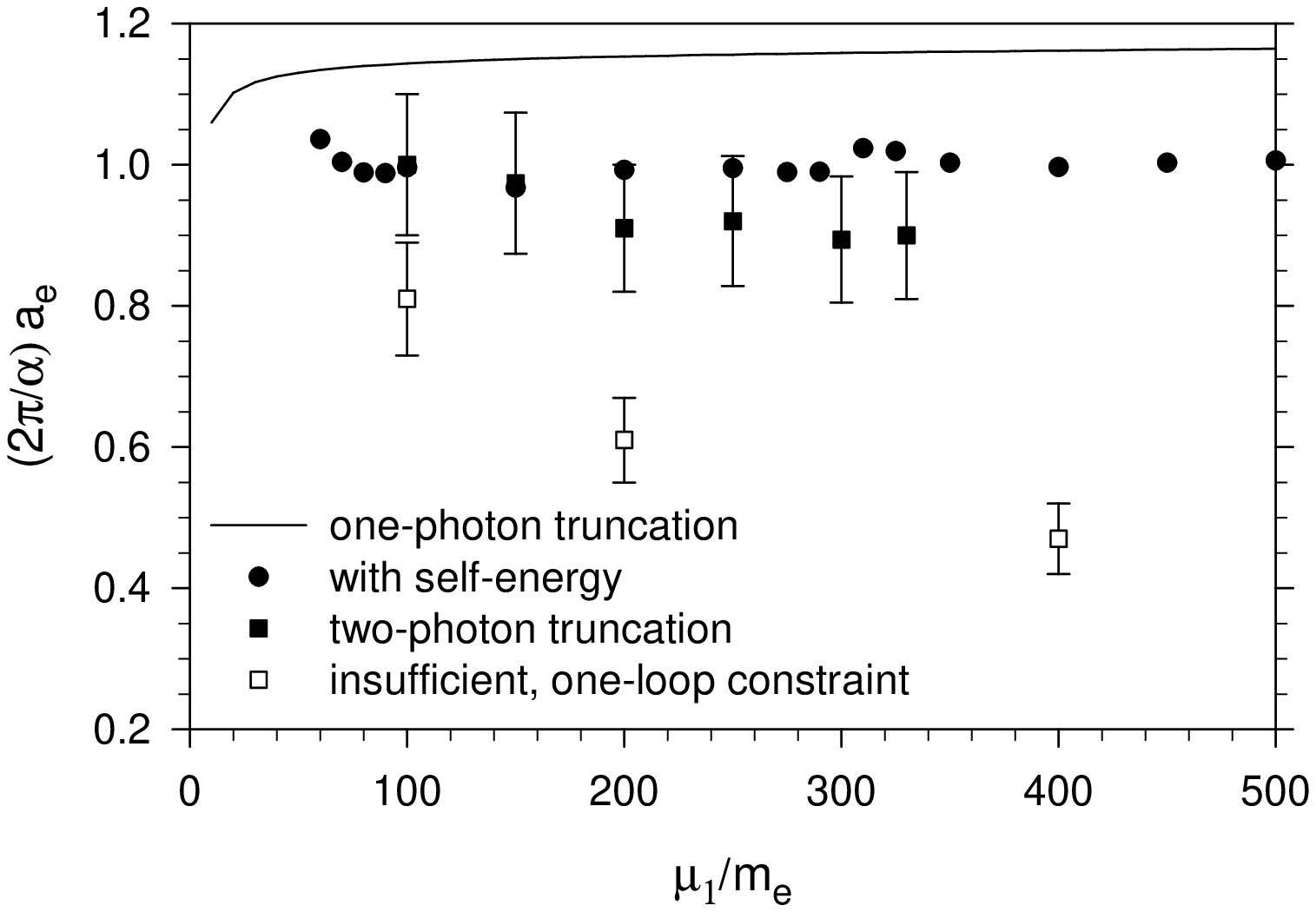}}
\caption{\label{fig:twophoton} The anomalous moment of the electron in
units of the Schwinger term ($\alpha/2\pi$) plotted versus
the PV photon mass, $\mu_1$, with the second PV photon mass, 
$\mu_2$, set to $\sqrt{2}\mu_1$
and the PV electron mass $m_1$ equal to $2\cdot10^4\,m_e$.
The solid squares are the result of the full two-photon truncation
with the correct, nonperturbative chiral constraint~\protect\cite{TwoPhotonQED}.
The open squares come from use of a perturbative, one-loop constraint.
Results for the one-photon truncation~\protect\cite{ChiralLimit}
(solid line) and the one-photon truncation with the
two-photon self-energy contribution~\protect\cite{SecDep} (filled circles)
are included for comparison.
The resolutions used for the two-photon results are $K=50$ to 150,
combined with extrapolation to $K=\infty$, and $N_\perp=20$.}
\end{figure}

\section{Results}

From the solutions to the eigenvalue problems, we compute the 
anomalous moment at fixed PV masses and fixed numerical
resolution.  We then study the behavior first as a function
of the numerical resolution, which requires extrapolation,
and then as a function of PV masses.  The numerical
resolution is marked by two parameters, $K$ and $N_\perp$, which
control the number of quadrature points used in the longitudinal
and transverse directions.  The numerical convergence
and extrapolation are illustrated in \cite{TwoPhotonQED}.

The results of the extrapolations are plotted in Fig.~\ref{fig:twophoton}.
Each value is close to the standard Schwinger result of
$\alpha/2\pi$ and independent of $\mu_1$, to within numerical error.
The results with only the two-photon self-energy contribution
are actually better than the full two-photon results.  This
discrepancy should be due to the absence of electron-positron
contributions, which are of the same order in $\alpha$ as
the two-photon contributions; without the electron-positron
contributions, we lack the cancellations that typically take place
between contributions of the same order.

If we retain only the
self-energy contributions from the two-photon intermediate states,
the equations for the two-body wave functions become much simpler, and
the coupled integral equations can be reduced to the one-electron
sector.  There, they can be solved analytically, except for the
calculation of certain integrals~\cite{SecDep}.
We see that the inclusion of the self-energy contribution
is a significant improvement over the one-photon truncation.
Thus, we expect that inclusion of the three-photon self-energy
will improve the two-photon results.

Figure~\ref{fig:twophoton} also includes results obtained for the
two-photon truncation when only the one-loop chiral constraint
is satisfied.  Without the full
nonperturbative constraint, the results are very sensitive
to the PV photon mass $\mu_1$.  This behavior repeats the pattern observed
in \cite{ChiralLimit} for a one-photon truncation without the
corresponding one-loop constraint.  The resulting $\mu_1$
dependence is illustrated in Fig.~2 of \cite{ChiralLimit}.  Thus,
a successful calculation requires that the symmetry of the chiral
limit be maintained.

\acknowledgments
The work reported here was done in collaboration with 
J.R. Hiller
and supported in part by the Minnesota Supercomputing Institute.

\end{document}